\documentclass[11pt]{article}

\usepackage[utf8]{inputenc}
\usepackage[T1]{fontenc}
\usepackage{graphicx}
\usepackage{amssymb}
\usepackage{xcolor}
\usepackage{siunitx}
\usepackage{amsmath}
\usepackage{hyperref}
\usepackage{tabularx}
\usepackage{booktabs}

\usepackage[numbers,sort&compress]{natbib}

\title{On the Physical Nature of the Scalar Mode Mass in the Jordan Frame of Metric $f(R)$ Gravity}

\author{
Giovanni Montani$^{1,2}$ \and
Andrea Valletta$^{1,3}$\thanks{Corresponding author: andrea.valletta@uniroma1.it}
}

\date{}

\begin{document}

\maketitle

\begin{center}
{\small
$^{1}$ Physics Department, Sapienza University of Rome, P.le A. Moro 5, 00185 Rome, Italy \\
$^{2}$ Nuclear Department, ENEA, C.R. Frascati, Via E. Fermi 45, 00044 Frascati, Italy \\
$^{3}$ INFN, Sezione di Roma, P.le A. Moro 5, 00185 Rome, Italy
}
\end{center}

\vspace{1em}

\begin{abstract}
We analyze the Taylor expansion of the metric $f(R)$ gravity in the Jordan frame around the General Relativity limit, expanding in the small deviation $(\phi-\phi_0)$ with $\phi_0=1$. By relating the scalar--tensor representation to the original $f(R)$ formulation, we derive constraints on the expansion parameters from the observed value of the present-day $\Lambda$CDM deceleration parameter and from cosmological bounds on the variation of Newton's constant. We show that these requirements imply that the scalar degree of freedom must have a mass exceeding the Hubble scale by several orders of magnitude. This result challenges the common assumption that the scalar mode can drive cosmological dynamics with a mass of order of the Hubble constant $H_0$. We provide a dynamical interpretation of this hierarchy by emphasizing that a proper definition of the scalar mass, in a field-theoretical sense, requires an adiabatic separation between background evolution and perturbations, which naturally leads to a super-Hubble mass scale.
\end{abstract}

\vspace{1em}

\noindent\textbf{Keywords:} $f(R)$ gravity; Jordan frame; scalar field mass; Hubble tension

\section{Introduction} 

A natural extension of General Relativity is provided by the so-called $f(R)$ gravity theories
~\cite{SotiriouFaraoni2010, capozziello2006observational, Odintsov:2020qzd, Odintsov:2023cli}. Such models are constructed by a modification of the Einstein--Hilbert action, in~which the Ricci scalar $R$ is replaced by a generic function $f(R)$. This procedure leads to a rich phenomenology that is absent in General Relativity~\cite{montani2024metric, Valletta:2025bgu, schiavone2023f, Montani2025, olmo2005gravity}.
It is well known that $f(R)$ gravity theories are equivalent to scalar-tensor theories, in~which the scalar field couples either to the metric tensor (Jordan frame) or to the matter sector (Einstein frame). In~this work, we will use only metric $f(R)$ gravity in the Jordan frame, in~which the self-interacting scalar field is non-minimally coupled to the metric tensor, as~described in~\cite{SotiriouFaraoni2010,capozziello2011extended}.

Chronologically, metric $f(R)$ gravity was first investigated in the context of early universe cosmology, in~particular to address the Big Bang singularity~\cite{starobinsky1980new}, and~later applied to spherically symmetric black hole spacetimes to regularize central curvature singularities~\cite{elizalde2011nonsingular}.

Recently, after~the discovery of the late-Universe acceleration~\cite{riess1998observational, perlmutter1999measurements}, $f(R)$ gravity has gained increasing interest, as~these theories can effectively reproduce the phenomenology of a negative-pressure fluid through a purely geometric modification of gravity~\cite{NojiriOdintsovPhysRep}. For~the metric-affine formulation see also~\cite{olmo2011palatini,bombacigno2021big}.

A simple and well-known example of an $f(R)$ theory is 
the so-called $R^2$ gravity~\mbox{\cite{capozziello2011extended, Hell:2023mph, Hell:2025wha, Hell:2026blj}}, originally introduced in~\cite{starobinsky1980new, starobinsky1987general}.
In particular, if~we also add a cosmological constant term in the $R^2$ gravity action, we are de facto

looking at the first three terms 
of a Taylor expansion of the generic function $f(R)$ at sufficiently 
low curvatures. 
Two fundamental requirements naturally arise when modifying General~Relativity: 
\begin{itemize}
    \item Any correction to the standard gravity theory must be in agreement, within~experimental uncertainties, with~the current tests of GR, the~most stringent among which are the Solar System tests~\cite{barrow1983stability}.
    \item Any new theory of gravity has to account for the late-time accelerated expansion of the Universe, which takes place at large scales. To~do so, the~theory introduces corrections that become important at low values of $R$. See, for~example, the~Hu--Sawicki model~\cite{hu2007models}.
\end{itemize}

A further challenge in $f(R)$ gravity is the so-called ``Hubble tension'', i.e.,~the $4\sigma$ discrepancy in the detections of the Hubble constant value 
from different sources. On the~one hand, {the Supernova H0 for the Equation of State (S$H_0$ES) collaboration~\cite{Riess2022} that measures $H_0$ from the supernova data at low redshift $z$ ($z\sim 0.01$--$0.1$); while on the other hand the Planck satellite collaboration~\cite{Planck2018}} infers $H_0$ from the CMB measurements at $z\sim1100$. For~a comprehensive review of the various determinations of the Hubble constant, see~\cite{di2021realm,Abdalla:2022yfr}. 
Many theoretical models have been proposed to alleviate the Hubble tension, with~promising results, see, for~instance,~\cite{NojiriOdintsovPhysRep, schiavone2023f, montani2024metric, montani2025decay, montani2024modified, Valletta:2025bgu}. See also~\cite{dainotti2021hubble, Dainotti2022b, Odintsov:2020qzd, Odintsov:2023cli, Nojiri:2022ski, desimone2024doubletcosmologicalmodelschallenge, Dainotti:2024aha}.

The capacity of metric $f(R)$ gravity to account for both the Universe acceleration and the Hubble tension does not come without limitations. To~satisfy Solar System tests, the~mass of the scalar degree of freedom must be sufficiently large to suppress fifth-force effects. Moreover, different experiments~\cite{li2018measurements, gundlach2000measurement} impose strong local constraints on the value of the Newton constant $G$, and~hence on the coefficient of the term proportional to $R$ in the Taylor series expansion of $f(R)$ at low curvatures $R_0$. In~particular, $|\frac{\Delta G}{G}|\le 10^{-5}$~\cite{li2018measurements, gundlach2000measurement} ($\Delta G$ is the maximum variation of the Newton constant according to the local tests).
Although we assume $G$ to be constant, hence independent from the place in which it is measured, we prefer to use cosmological constraints on it as a consequence of the fact that we are in cosmological regimes. Hence, for~the constraint on the Newtonian constant, we use the results from~\cite{wang2020constraints}, according to which $\frac{\Delta G}{G}\le 3\%$. This assumption fixes $|f'(R_0)-1|\le3\cdot10^{-2}$. 
In Table~\ref{Tab_1} we summarize all the cosmological constraints we use in this~manuscript.

\begin{table}[htbp]
\centering
\caption{Cosmological constraints used in this work.}
\begin{tabularx}{0.95\textwidth}{l c c}
\toprule
\textbf{Quantity} & \textbf{Constraint} & \textbf{Reference} \\
\midrule
Hubble constant $H_0$ & $\sim 70\ \mathrm{km\,s^{-1}\,Mpc^{-1}}$ & \cite{Riess2022,Planck2018} \\
Deceleration parameter $q_0$ & $\simeq -0.55$ & \cite{weinberg2008cosmology} \\
Variation of Newton constant $\Delta G/G$ & $\le 3\%$ & \cite{wang2020constraints} \\
\bottomrule
\end{tabularx}
\label{Tab_1}
\end{table}

To simultaneously satisfy local (Solar System) constraints and produce effects at cosmological scales, a~possible explanation is that the scalar field undergoes a screening mechanism.  
Among those proposed, the~so-called ``chameleon behavior''~\cite{khoury2004chameleon, khoury2004chameleon2, carroll1998quintessence, quiros2015chameleon, brax2008f, PhysRevD.77.107501, LIU2018286, 10.1093/mnras/sty1822} has received considerable attention. In~this mechanism, the~mass of the scalar mode depends on the local energy density: in other words, the~scalar field mass decreases as the spatial scale increases. 
Consequently, within~the high-density environment of the Solar System, the~Compton wavelength of the scalar field is short, in~such a way that the field becomes effectively non-interacting. On~the contrary, in~the low-density cosmological environment, the~scalar field acquires a mass of order $H_0$ (i.e. a Compton scale of the Boson field of the order of the Hubble scale), allowing it to drive the accelerated expansion of the~Universe.

Although such a formulation could address several theoretical issues, a~complete physical and geometrical model that naturally leads to the chameleon mechanism is still lacking (see~\cite{amendola2007conditions}).

{The relation between the scalar field mass, local gravity constraints and cosmological dynamics in $f(R)$ gravity has been widely investigated in the literature. 
In particular, several studies have analyzed the compatibility between modified gravity models and Solar System experiments by invoking screening mechanisms such as the chameleon effect. 
For instance, the~analysis in~\cite{10.1093/mnras/sty1822, PhysRevD.77.107501, LIU2018286} explores specific functional forms of $f(R)$ (such as Hu--Sawicki or power-law models) and constrains their parameters using a combination of local gravity tests, cosmological evolution and astrophysical observations.}

{In these works the recovery of General Relativity typically relies on the presence of a screening mechanism that suppresses the scalar interaction in high-density environments. 
However, the~scalar mass itself is not derived from purely cosmological consistency conditions, but~is instead determined by the environmental dependence introduced by the screening mechanism.}

{From the point of view of the underlying theory, however, this mechanism effectively introduces a coupling between the scalar field dynamics and the environmental matter density~\cite{amendola2007conditions}. 
Such a coupling weakens the purely geometrical interpretation of the $f(R)$ Lagrangian, since the scalar sector acquires an effective interaction with the matter content of spacetime. In~this respect, the~result obtained in the present work is related to, but~conceptually different from, the~chameleon scenario. 
In fact, our constraint on the scalar field mass is derived entirely from cosmological considerations, namely from the observed value of the deceleration parameter and from bounds on the variation of Newton's constant inferred from cosmological observations. 
Therefore, the~requirement that the scalar mass must largely exceed the Hubble scale emerges independently of Solar System tests and provides a direct consistency condition on the cosmological dynamics of analytic $f(R)$ models.}

In this paper, we focus our attention on a related question: is it possible to reconcile the {deceleration parameter $q_0$} of the $\Lambda$CDM model with a scalar field mass of order $H_0$, while simultaneously satisfying the experimental constraints on the variation of Newton's constant $\Delta G$?

Our analysis shows that, in~order to satisfy these constraints and remain consistent with the observed value of the deceleration parameter, the~scalar mode mass must exceed the Hubble scale by roughly three (five if we consider local measurements of the Newtonian constant instead of cosmological ones) orders of magnitude. Furthermore, we provide a physical explanation for this result that is consistent with the standard interpretation of mass terms in classical field theory~\cite{mandl2013quantum}.

\section{Scalar Field Dynamics and Constraints}
\label{Sec. teoria}

{We begin by recalling that metric $f(R)$ gravity can be written in a scalar--tensor form in the Jordan frame. The~action takes the form
\begin{equation}
\mathcal{S}=\frac{1}{2\chi}\int d^4x\,\sqrt{-g}\left[\phi R - V(\phi)\right] + S_m ,
\end{equation}
where $\chi$ denotes the Einstein constant, $g$ is the determinant of the metric tensor, $R$ is the Ricci scalar, and~$S_m$ represents the matter action. The~scalar field $\phi$ corresponds to the additional dynamical degree of freedom introduced by the modification of gravity and is defined through
\begin{equation}
\phi \equiv \frac{df}{dR}.
\end{equation}}

A basic viability requirement for a given $f(R)$ model is that the scalar mode possesses a positive effective mass. In~the scalar–tensor representation this mass is defined as
\begin{equation}
m_\phi^2 \equiv \frac{1}{3}\left(\phi\frac{d^2V}{d\phi^2}-\frac{dV}{d\phi}\right),
\end{equation}
which guarantees the stability of the scalar degree of~freedom.

{Variation of the action with respect to the metric tensor provides the modified Einstein equations. Taking their trace and using the relation $R=dV/d\phi$, one obtains a Klein–Gordon-like equation for the scalar field. In~the presence of matter sources, the~trace of the energy–momentum tensor contributes as a source term for the scalar field dynamics.}

We consider a spatially flat Friedmann–Lemaître–Robertson–Walker (FLRW) spacetime with~a line element
~\cite{friedman1922krummung, lemaitre1927univers, robertson1935kinematics, walker1937milne}, according to the Planck measurements~\cite{efstathiou2020evidence},
\begin{equation}
ds^2 = -dt^2 + a^2(t)\,\delta_{ij}dx^i dx^j \, ,
\end{equation}
where \( a(t) \) is the cosmic scale factor and \( t \) denotes synchronous~time.

In this background, the~modified Friedmann equations in metric \( f(R) \) gravity, as~viewed in the so-called ``Jordan frame''~\cite{SotiriouFaraoni2010}, read as follows:
\begin{equation}
	H^2 \equiv \left( \frac{\dot{a}}{a}\right)^2 = \frac{1}{\phi}\left( 
	\frac{\chi}{3}\rho_m - H \dot{\phi} + \frac{V(\phi )}{6}\right)
	\, 
	\label{frs1}
\end{equation}
and
\begin{equation}
	\frac{dV}{d\phi} = 12H^2 + 6\dot{H}
	\,, 
	\label{frs2}
\end{equation}
where $\chi$ is the Einstein constant (we are in $c=1$ units),  the~dot indicates differentiation with respect to $t$, and $\phi$ denotes the non-minimally coupled scalar field emerging in the Jordan frame, whose self-interaction is described by the potential term $V(\phi )$.
Finally, the~quantity $\rho_m$ stands for the energy density of the (dark and
baryonic) matter component of the~Universe.

The first of these equations is the 
generalized Friedmann equation, while 
the second one is obtained by varying 
the gravitational action with respect to the scalar field $\phi$. 

Now, passing to the time variable 
$x=\ln(1+z)=\ln(1/a)$ 
($z$ denoting the redshift and we 
set the present-day value of the cosmic scale factor to unity, i.e.,~$a_0\equiv a(z=0)=1$), the~two equations above can be rewritten as
\begin{equation}
	H^2=\frac{1}{\phi - \frac{d\phi}{dx}}\left( \frac{\chi}{3}\rho_m(x) 
	+ \frac{V(\phi(x))}{6}\right)
	\, 
	\label{frs3}
\end{equation}
and
\begin{equation}
	\frac{dV}{d\phi} = 12 H^2 - 
	3\frac{dH^2}{dx}
	\,, 
	\label{frs4}
\end{equation}
respectively.
Above, we made use of the relation 
$\dot{(\dots)}= -Hd(\dots)/dx$.

{We now intend to study Equations~(\ref{frs3}) and (\ref{frs4}) very close to us, i.e.,~for $x\rightarrow 0$.
To this end, we perform a Taylor expansion of
the scalar field dependence for 
$\phi \simeq \phi_0\equiv \phi(x=0)$, so that}

{\begin{align}
V(\phi) \simeq {} &
V(\phi_0)
+ \left( \frac{dV}{d\phi} \right)_{\phi=\phi_0}(\phi - \phi_0)
+
\\
&\frac{1}{2}
\left( \frac{d^2 V}{d\phi^2} \right)_{\phi=\phi_0}
(\phi - \phi_0)^2
+ \ldots
\nonumber \\
= {} &
V_0
+ \bar{V}(\phi - \phi_0)
+ \frac{1}{2} V^*(\phi - \phi_0)^2
+ \ldots
\label{frs5}
\end{align}}

{The relation between the potential 
$V(\phi )$ and the underlying $f(R)$ Lagrangian is dictated via the algorithm}

{\begin{equation}
	R = \frac{dV}{d\phi} \quad, \, f(R) \equiv \phi\frac{dV}{d\phi} - V(\phi )
	\,, 
	\label{frs6}
\end{equation}}
{where the function $\phi(R)$ in the right-hand-side of the second equation above is obtained by assuming the 
invertibility of $dV/d\phi$ in the first one. }

{By implementing Equation~(\ref{frs6}) 
to calculate the $f(R)$ expression, that is 
associated with the potential term 
in Equation~(\ref{frs5}), the~first derivative gives
\begin{equation}
\frac{dV}{d\phi} = \bar V + V^* (\phi - \phi_0) + \dots = R \,,
\end{equation}
which can be inverted at leading order to yield
\begin{equation}
\phi - \phi_0 \simeq \frac{R - \bar V}{V^*}.
\end{equation}}

{Plugging this into the definition
\begin{equation}
f(R) = \phi R - V(\phi),
\end{equation}
we obtain the following step by step:
\begin{align}
f(R) &\simeq (\phi_0 + (\phi - \phi_0)) R - \left[ V_0 + \bar V (\phi - \phi_0) + \frac{1}{2} V^* (\phi - \phi_0)^2 \right] \\
&=  - V_0 + \left(\phi_0 - \frac{\bar V}{V^*}\right) R + \frac{1}{2 V^*} R^2 + \frac{\bar V^2}{2 V^*} + \dots
\label{frs7}
\end{align}}

{This result directly follows from the Taylor expansion around the General Relativity limit $\phi_0=1$ and illustrates how the parameters $V_0$, $\bar V$, and~$V^*$ control the deviation from standard GR.}

{In order to restore the Einsteinian gravity in the low-curvature limit, we require the following conditions on the expansion parameters:
\begin{equation}
\phi_0 = 1, \quad V_0 \equiv \bar V \equiv 0, \quad V^* \gg H_0^2,
\end{equation}
where $\phi_0=1$ ensures that the effective Newton constant matches its standard value, and~$V_0$ and $\bar V$ vanish 
to avoid introducing a cosmological term beyond the one we explicitly include if desired. The~condition $V^* \gg H_0^2$ guarantees a sufficiently heavy scalar mode to satisfy local and cosmological constraints. These conditions, applied to Equation~(\ref{frs7}), restore GR (i.e., $f(R)=R$).}

However, we can also retain a 
cosmological constant term by~defining $V_0 \equiv 2\chi \rho_{\Lambda}$ ($\rho_{\Lambda}$ being the vacuum energy density). 

Evaluating Equation~(\ref{frs4}) 
in $x=0$ (i.e., $\phi=\phi_0$) 
according to the expansion in Equation~(\ref{frs5}), we obtain
\begin{equation}
	\bar{U}\equiv \frac{\bar{V}}{6H_0^2} = 2 - \frac{1}{2H_0^2}
	\left(\frac{dH^2}{dx}\right)_{x=0} = 1 - q_0
	\,, 
    \label{frs9}
\end{equation}
where we introduced the deceleration 
parameter $q_0$ of the Universe, 
defined as in~\cite{weinberg2008cosmology}:
\begin{equation}
	q_0 =  -1 + \frac{1}{H_0}
    \left(\frac{dH}{dx}\right)_{x=0}
	\, .
	\label{frs10}
\end{equation}

Since we have $q_0\simeq -0.55$ for~the $\Lambda$CDM model, we should have
\begin{equation}
    \bar{V}\simeq 9.3\times H_0^2\simeq 
10^{-52}\,\mathrm{m^{-2}}
\label{valoreVbar}
\end{equation}
 at cosmological scales. The~situation does not change for models that slightly deviate from the standard cosmological one, such as those in which the 
Hubble tension~\cite{di2025cosmoverse, di2021realm} is addressed through small changes in the matter~\cite{navonea2025creation, fazzari2025effective} or gravity sector of the~theory. 

{Now, from~the measurements of the Newton constant~\cite{wang2020constraints}, 
the coefficient multiplying the Ricci scalar in the effective Einstein--Hilbert term 
cannot deviate significantly from unity. In~metric $f(R)$ gravity this coefficient is 
given by $f'(R)$, which plays the role of an effective gravitational coupling. 
Cosmological observations constrain its present value to satisfy}

{\begin{equation}
|f'(R_0)-1| \lesssim 3 \times 10^{-2}.
\end{equation}}

{Using the expansion obtained in Equation~(\ref{frs7}) and evaluating it around the present epoch 
$\phi_0 \equiv 1$, we have}

{\begin{equation}
f'(R_0) = \phi_0 - \frac{\bar V}{V^*}.
\end{equation}}

{Therefore the observational constraint on the variation of Newton's constant implies}

{\begin{equation}
\left|\frac{\bar V}{V^*}\right| \lesssim 3\times10^{-2}.
\end{equation}}

{Since Equation~(\ref{valoreVbar}) gives $\bar V \simeq 10^{-52}\,\mathrm{m^{-2}}$, the~previous 
relation immediately yields the bound} 
{\begin{equation}
V^* \gtrsim 10^{-50}\,\mathrm{m^{-2}} .
\end{equation}

The scalar field mass, as~defined in~\cite{olmo2007violation}, takes, at $x=0$, the form}

{\begin{equation}
m_{\phi}^2 = \frac{1}{3}(V^*-\bar V)
\simeq \frac{V^*}{3}
>10^{-50}\,\mathrm{m^{-2}}
\gg H_0^2 .
\label{massa}
\end{equation}}

{The result obtained above follows from the combination of three independent ingredients:}

{\begin{itemize}
\item[(i)] The observed value of the present-day deceleration parameter 
$q_0 \simeq -0.55$, which fixes the linear coefficient of the potential expansion 
through Equation~(\ref{frs9}), yielding $\bar V \sim 10^{-52}\,\mathrm{m^{-2}}$.

\item[(ii)] The observational constraint on the variation of the Newton constant, 
which requires $|f'(R_0)-1| \lesssim 3\times10^{-2}$ and therefore implies 
$V^* \gtrsim 10^{-50}\,\mathrm{m^{-2}}$ from Equation~(\ref{frs7}).

\item[(iii)] The definition of the scalar-field mass in metric $f(R)$ gravity,
$m_\phi^2=\frac{1}{3}(V^*-\bar V)$.
\end{itemize}}

Combining these relations leads~to

\[
m_\phi^2 \gtrsim 10^{-50}\,\mathrm{m^{-2}} \gg H_0^2 ,
\]
which implies that the quadratic scalar field mass is roughly $10^2$ times larger than the 
Hubble~scale.

This result is in apparent contradiction with the common expectation that the 
non-minimally coupled scalar field driving cosmological dynamics should have a 
Compton wavelength of the order of the Hubble length~\cite{khoury2004chameleon, quiros2015chameleon, farajollahi2010cosmic}.

Actually, many metric $f(R)$ models, 
providing valuable cosmological phenomenology~\cite{hu2007models, schiavone2023f, montani2024metric, Valletta:2025bgu}, work only if the value of the 
scalar field mass is much smaller 
than required by Solar System tests~\cite{capozziello2011extended}.

{However, to~address the present discrepancy, we briefly reconsider the derivation 
of the scalar-field mass in metric $f(R)$ gravity, following the procedure discussed 
in~\cite{olmo2007violation}.}

{On a flat isotropic cosmological background, the~Klein–Gordon equation of the 
non-minimally coupled scalar field emerging in the Jordan frame reads as}

{\begin{equation}
3\ddot{\phi} + 9H\dot{\phi} + 2V - \phi \frac{dV}{d\phi} = \chi \rho_m .
\end{equation}}

{To identify the mass of the scalar degree of freedom, one expands the field 
around a background cosmological solution $\phi_0(t)$ and introduces a small 
perturbation $\delta\phi(t)$,}

{\begin{equation}
\phi(t)=\phi_0(t)+\delta\phi(t),
\end{equation}}
{with $|\delta\phi|\ll\phi_0$. Substituting this decomposition into the equation 
above and expanding to first order in $\delta\phi$, one obtains two equations: 
a background equation for $\phi_0$ and a linearized equation governing the 
perturbation.}

{The background dynamics is described by}

{\begin{equation}
3\ddot{\phi}_0 + 9H\dot{\phi}_0 + 2V(\phi_0) 
- \phi_0\left(\frac{dV}{d\phi}\right)_{\phi_0}
= \chi\rho_m ,
\end{equation}}
{while the perturbation satisfies}

{\begin{equation}
\ddot{\delta\phi}+3H\dot{\delta\phi}+m_\phi^2\,\delta\phi=0 ,
\label{pert}
\end{equation}}
{which has the form of a Klein–Gordon equation for a massive scalar mode. 
The corresponding mass term is therefore}

{\begin{equation}
m_\phi^2=
\frac{1}{3}\left[
\phi_0\left(\frac{d^2V}{d\phi^2}\right)_{\phi_0}
-
\left(\frac{dV}{d\phi}\right)_{\phi_0}
\right].
\label{once}
\end{equation}}

{The expression above, for~the expansion in Equation~(\ref{frs5}) and $\phi_0=1$ 
yields Equation~(\ref{massa}).}

{Another consequence can be obtained taking the second derivative of the potential with respect to the field (from Equation~(\ref{frs2})):}

{\begin{equation}
    \frac{d^2V}{d\phi^2}=24H\frac{dH}{d\phi}-6\frac{dH}{d\phi}-6H\frac{d^2H}{d\phi dx}=\frac{6}{\frac
{d\phi}{dx}}\left[4H\frac{dH}{dx}-\left(\frac{dH}{dx}\right)^2-H\frac{d^2H}{dx^2}\right].
\label{derivata seconda potenziale}
\end{equation}}

{Using now the well-known relations}

{\begin{equation}
    \frac{dH}{dz}=\frac{H(1+q)}{1+z},
    \label{eq:1}
\end{equation}}
{\begin{equation}
    \frac{d^2H}{dz^2}=\frac{H}{(1+z)^2}(J-q^2);
    \label{eq.2}
\end{equation}}
{and inserting (\ref{derivata seconda potenziale}) in the inequality of Equation~(\ref{massa}),}

{\begin{equation}
    \frac{V^*}{3}\equiv \frac{1}{3}\frac{d^2 V}{d\phi^2}|_{x=0}>10^{-50}\,\mathrm{m^{-2}}
\end{equation}}
{produces the following result:}

{\begin{equation}
    \frac{d\phi}{dx}|_{x=0}<2\cdot10^{50}(3+2q_0-J_0)H_0^2\,\mathrm{m^{2}}.
    \label{disequazione}
\end{equation}}

{Now, using the cosmographic parameters from the $\Lambda CDM$ model ($q_0=-0.55$, $J_0=1$), the~inequality (\ref{disequazione}) becomes}

{\begin{equation}
    \frac{d\phi}{dx}|_{x=0}<10^{-2}.
    \label{finale}
\end{equation}}

{This result has a profound consequence on the value of the mass of the scalar field, in~particular on its divergence in $z=0$, as~it can be seen in~\cite{Valletta:2025bgu}.}

An important point here is that the 
expression of the mass in Equation~(\ref{once}) makes sense in Equation~(\ref{pert}) only if the perturbation $\delta \phi$ is varying on a much faster time scale with respect to that one of the 
background variation (see. for reference,~\cite{olmo2005gravity}). In~fact, in~this case, the~value of the mass, at~a given instant, can be regarded as frozen 
and Equation~(\ref{pert}) becomes a real 
Klein--Gordon equation on an adiabatically varying cosmological~background.

In this respect, our estimates, based on cosmological bounds on the variation of Newton's constant, 
suggest that the time scale over 
which the massive scalar field 
formulation evolves is, at~least, 
$10^2$ times shorter than the Hubble time and that it seems meaningless 
to discuss the non-minimally 
coupled scalar field mass on 
a much larger scale. 
In other words, we infer that, 
at the Hubble scale, $\phi$ is just 
a non-trivial scalar mode, coupled to gravity and no real concept of 
physical mass can be attributed to it. 
Different would be the story for 
a fast scalar fluctuation, which, instead, behaves, on~a linear approach, 
as a massive mode. This consideration acquires particular interest in the 
spirit of the analysis in~\cite{montani2025decay2}, where the possible decaying 
process of a fast massive fluctuation 
of $\phi$ into dark matter particles has been studied, 
although the formulation was limited by 
including a phenomenological decaying rate (see also~\cite{montani2025decay}).

{As an illustrative example, let us consider the logarithmic model proposed in~\cite{kruglov2023logarithmic}, defined by the modified $f(R)$ Lagrangian}

{\begin{equation}
f(R) = -\frac{1}{\beta}\ln(1-\beta R) ,
\label{log1}
\end{equation}}
{where $\beta$ is a free parameter of the model with the dimension of a squared length. Local gravity constraints require the bound $\beta < 6 \times 10^{-6}\,\mathrm{cm}^2$.}

{Expanding Equation~(\ref{log1}) in the regime $\beta R \ll 1$, one obtains}

{\begin{equation}
f(R) \simeq R + \frac{\beta}{2}R^2 + \mathcal{O}(R^3).
\end{equation}}

{Comparing this expansion with the general form obtained from the Taylor expansion of the potential in Equation~(\ref{frs7}), we immediately obtain the relations}

{\begin{equation}
\bar{V} \equiv 0, \qquad V_0 = V^* = \frac{1}{\beta}.
\label{log2}
\end{equation}}

{Using Equation~(\ref{frs9}), which relates $\bar{V}$ to the present-day value of the deceleration parameter, we see that the condition $\bar{V}=0$ implies}

{\begin{equation}
q_0 = 1 .
\end{equation}}

{This result clearly contradicts the observational value $q_0 \simeq -0.55$. Therefore, the~logarithmic model cannot reproduce the present cosmic acceleration within the perturbative expansion around the General Relativity limit discussed above. The~accelerated expansion found in~\cite{kruglov2023logarithmic} must then arise as a genuinely non-perturbative effect, in~which contributions from all orders of the expansion become relevant.}

{The example of a logarithmic Lagrangian, as~that one discussed above, is paradigmatic of the applicability of our constraint to extract physical information on the cosmological dynamics predicted by a given model. However, it is worth stressing that this situation does not apply to non-analytic $f(R)$ Lagrangians, i.e.,~theories that cannot be expanded in Taylor series around $R=0$. In~such cases the perturbative procedure adopted in the present work is not applicable and the corresponding cosmological dynamics must be studied using different techniques~\cite{lecian2009implications, fiorucci2014non, capozziello2011cosmological, capozziello2013cosmological}.}

{Similarly, the~present analysis is not directly relevant for those $f(R)$ models whose structure implies $V^* \equiv 0$ in the scalar–tensor representation, corresponding to a vanishing scalar field mass. This situation occurs, for~instance, in~models such as the Hu–Sawicki construction~\cite{hu2007models}, where the recovery of General Relativity relies on screening mechanisms rather than on a perturbative expansion around the GR limit.}

{Furthermore, from~the general expression of the scalar field mass given in Equation~(\ref{once}), we obtain}

{\begin{equation}
m_\phi^2 = \frac{1}{3\beta}.
\end{equation}}

{If we require the adiabaticity condition discussed above, namely that the scalar perturbation evolves on a timescale much shorter than the cosmological background evolution, we must impose}

{\begin{equation}
\beta \ll H_0^{-2}.
\end{equation}}

{This condition is fully consistent with the local experimental bound on $\beta$ and confirms that the scalar mode behaves as a very massive field compared to the Hubble scale.}

\section{Conclusions}

In this work we have analyzed the Taylor expansion of metric $f(R)$ gravity in the Jordan frame around the General Relativity limit at the present cosmological epoch. 
Assuming that General Relativity must be recovered as the low-curvature limit of the theory, we expanded the scalar potential around $\phi_0 = 1$ and imposed two independent cosmological constraints: the observed value of the present-day deceleration parameter and the cosmological bound on the variation of Newton's~constant.

The combination of these requirements leads to a non-trivial consistency condition on the expansion parameters of the theory. In~particular, we have shown that the scalar degree of freedom necessarily acquires a mass much larger than the Hubble scale. Even adopting conservative cosmological bounds on $\Delta G/G$, the~resulting scalar mass squared exceeds $H_0^2$ by several orders of magnitude. If~one were to impose the more stringent Solar System limits on the variability of Newton's constant, the~hierarchy would become even stronger, yielding $m_\phi^2 > 10^5 H_0^2$.

It is important to stress the domain of validity of this analysis. Our result applies to analytic $f(R)$ models that admit General Relativity as an expansion limit around $\phi_0 = 1$. Well-known constructions such as the Hu–Sawicki model or other non-analytic formulations do not reproduce General Relativity through a regular Taylor expansion around this point. In~those scenarios, the~recovery of the GR limit typically relies on additional dynamical mechanisms, such as chameleon-type screening, which are introduced to ensure compatibility with local gravity~tests.

Therefore, our conclusion does not directly constrain such non-analytic models (i.e., models that cannot be Taylor expanded near $\phi-\phi_0$). Rather, it shows that within analytic metric $f(R)$ theories admitting GR as a perturbative limit at the present epoch, observational consistency alone forces a super-Hubble scalar mass~scale. 

Finally, we have emphasized that the notion of scalar mass in a cosmological background requires an adiabatic separation between background evolution and perturbations. When this condition is properly taken into account, the~emergence of a super-Hubble mass scale appears as a natural consequence of the consistency requirements imposed by~observations.

{A key novel aspect of our analysis is that, unlike the chameleon mechanism, the~requirement for the scalar field to have a large mass emerges solely from cosmological constraints, independent of local Solar System tests. This shows that the mass hierarchy is a consequence of the large-scale cosmological environment and not of local density effects, providing a new perspective on scalar field dynamics. This finding strengthens our results, highlighting that the super-Hubble scalar mass is an intrinsic cosmological feature rather than a consequence of screening mechanisms.}

\vspace{6pt}

\noindent\textbf{Author Contributions.}\\
\textbf{Giovanni Montani}: Conceptualization, Methodology, Formal analysis, Supervision, Writing—Original Draft, Writing—Review and Editing.  \\
\textbf{Andrea Valletta}: Conceptualization, Methodology, Formal Analysis, Visualization, Writing—Original Draft, Writing—Review and Editing.  
All authors have read and agreed to the published version of the manuscript.

\vspace{0.8em}

\noindent\textbf{Funding.}
This research received no external funding.

\vspace{0.8em}

\noindent\textbf{Data Availability Statement.}
The original contributions presented in this study are included in the article. Further inquiries can be directed to the corresponding authors.

\vspace{0.8em}

\noindent\textbf{Acknowledgments.}
We would like to thank Valerio Mangiapane for an interesting discussion about the inequality (\ref{finale}).

\vspace{0.8em}

\noindent\textbf{Conflicts of Interest.}
The authors declare no conflicts of interest.


\begin{thebibliography}{999}

\bibitem[Sotiriou and Faraoni(2010)]{SotiriouFaraoni2010}
Sotiriou, T.P.; Faraoni, V.
\newblock f(R) Theories of Gravity.
\newblock {\em Rev. Mod. Phys.} {\bf 2010}, {\em 82},~451--497.  
\newblock {\url{https://doi.org/10.1103/RevModPhys.82.451}}.

\bibitem[Capozziello et~al.(2006)Capozziello, Cardone, Elizalde, Nojiri, and Odintsov]{capozziello2006observational}
Capozziello, S.; Cardone, V.F.; Elizalde, E.; Nojiri, S.; Odintsov, S.D.
\newblock Observational constraints on dark energy with generalized equations of state.
\newblock {\em Phys. Rev. D—Part. Fields Gravit. Cosmol.} {\bf 2006}, {\em 73},~043512.

\bibitem[Odintsov et~al.(2021)Odintsov, S{\'a}ez-Chill{\'o}n~G{\'o}mez, and Sharov]{Odintsov:2020qzd}
Odintsov, S.D.; S{\'a}ez-Chill{\'o}n~G{\'o}mez, D.; Sharov, G.S.
\newblock {Analyzing the $H_0$ tension in $F(R)$ gravity models}.
\newblock {\em Nucl. Phys. B} {\bf 2021}, {\em 966},~115377. 
\newblock {\url{https://doi.org/10.1016/j.nuclphysb.2021.115377}}.

\bibitem[Odintsov et~al.(2023)Odintsov, Oikonomou, and Sharov]{Odintsov:2023cli}
Odintsov, S.D.; Oikonomou, V.K.; Sharov, G.S.
\newblock {Early dark energy with power-law F(R) gravity}.
\newblock {\em Phys. Lett. B} {\bf 2023}, {\em 843},~137988.  
\newblock {\url{https://doi.org/10.1016/j.physletb.2023.137988}}.

\bibitem[Montani et~al.(2024)Montani, De~Angelis, Bombacigno, and Carlevaro]{montani2024metric}
Montani, G.; De~Angelis, M.; Bombacigno, F.; Carlevaro, N.
\newblock Metric f (R) gravity with dynamical dark energy as a scenario for the Hubble tension.
\newblock {\em Mon. Not. R. Astron. Soc. Lett.} {\bf 2024}, {\em 527},~L156--L161.

\bibitem[Valletta et~al.(2025)Valletta, Montani, Dainotti, and Fazzari]{Valletta:2025bgu}
Valletta, A.; Montani, G.; Dainotti, M.G.; Fazzari, E.
\newblock {On the Metric $f(R)$ gravity Viability in Accounting for the Binned Supernovae Data}. \emph{arXiv
} {\bf 2025}, arXiv:2512.19568.

\bibitem[Schiavone et~al.(2023)Schiavone, Montani, and Bombacigno]{schiavone2023f}
Schiavone, T.; Montani, G.; Bombacigno, F.
\newblock f (R) gravity in the Jordan frame as a paradigm for the Hubble tension.
\newblock {\em Mon. Not. R. Astron. Soc. Lett.} {\bf 2023}, {\em 522},~L72--L77.

\bibitem[Montani et~al.(2025)Montani, Carlevaro, and Dainotti]{Montani2025}
Montani, G.; Carlevaro, N.; Dainotti, M.G.
\newblock Running Hubble Constant: Evolutionary Dark Energy.
\newblock {\em Phys. Dark Universe} {\bf 2025}, {\em 48}, 101847.
\newblock \href{https://doi.org/10.1016/j.dark.2025.101847}{https://doi.org/10.1016/j.dark.2025.101847}.



\bibitem[Olmo(2005)]{olmo2005gravity}
Olmo, G.J.
\newblock The Gravity Lagrangian according to solar system experiments.
\newblock {\em Phys. Rev. Lett.} {\bf 2005}, {\em 95},~261102.

\bibitem[Capozziello and De~Laurentis(2011)]{capozziello2011extended}
Capozziello, S.; De~Laurentis, M.
\newblock Extended theories of gravity.
\newblock {\em Phys. Rep.} {\bf 2011}, {\em 509},~167--321.

\bibitem[Starobinsky(1980)]{starobinsky1980new}
Starobinsky, A.A.
\newblock A new type of isotropic cosmological models without singularity.
\newblock {\em Phys. Lett. B} {\bf 1980}, {\em 91},~99--102.

\bibitem[Elizalde et~al.(2011)Elizalde, Nojiri, Odintsov, Sebastiani, and Zerbini]{elizalde2011nonsingular}
Elizalde, E.; Nojiri, S.; Odintsov, S.; Sebastiani, L.; Zerbini, S.
\newblock Nonsingular exponential gravity: A simple theory for early- and late-time accelerated expansion.
\newblock {\em Phys. Rev. D—Part. Fields Gravit. Cosmol.} {\bf 2011}, {\em 83},~086006.

\bibitem[Riess et~al.(1998)Riess, Filippenko, Challis, Clocchiatti, Diercks, Garnavich, Gilliland, Hogan, Jha, Kirshner, et~al.]{riess1998observational}
Riess, A.G.; Filippenko, A.V.; Challis, P.; Clocchiatti, A.; Diercks, A.; Garnavich, P.M.; Gilliland, R.L.; Hogan, C.J.; Jha, S.; Kirshner, R.P.;  et~al.
\newblock Observational evidence from supernovae for an accelerating universe and a cosmological constant.
\newblock {\em  Astron. J.} {\bf 1998}, {\em 116},~1009--1038.

\bibitem[Perlmutter et~al.(1999)Perlmutter, Aldering, Goldhaber, Knop, Nugent, Castro, Deustua, Fabbro, Goobar, Groom, et~al.]{perlmutter1999measurements}
Perlmutter, S.; Aldering, G.; Goldhaber, G.; Knop, R.A.; Nugent, P.; Castro, P.G.; Deustua, S.; Fabbro, S.; Goobar, A.; Groom, D.E.;  et~al.
\newblock Measurements of $\Omega$ and $\Lambda$ from 42 high-redshift supernovae.
\newblock {\em  Astrophys. J.} {\bf 1999}, {\em 517},~565--586.

\bibitem[Nojiri and Odintsov(2011)]{NojiriOdintsovPhysRep}
Nojiri, S.; Odintsov, S.D.
\newblock {Unified cosmic history in modified gravity: From F(R) theory to Lorentz non-invariant models}.
\newblock {\em Phys. Rept.} {\bf 2011}, {\em 505},~59--144.  
\newblock {\url{https://doi.org/10.1016/j.physrep.2011.04.001}}.

\bibitem[Olmo(2011)]{olmo2011palatini}
Olmo, G.J.
\newblock Palatini approach to modified gravity: F (R) theories and beyond.
\newblock {\em Int. J. Mod. Phys. D} {\bf 2011}, {\em 20},~413--462.

\bibitem[Bombacigno et~al.(2021)Bombacigno, Boudet, Olmo, and Montani]{bombacigno2021big}
Bombacigno, F.; Boudet, S.; Olmo, G.J.; Montani, G.
\newblock Big bounce and future time singularity resolution in Bianchi I cosmologies: The projective invariant Nieh-Yan case.
\newblock {\em Phys. Rev. D} {\bf 2021}, {\em 103},~124031.

\bibitem[Hell et~al.(2024)Hell, Lust, and Zoupanos]{Hell:2023mph}
Hell, A.; Lust, D.; Zoupanos, G.
\newblock {On the degrees of freedom of R$^{2}$ gravity in flat spacetime}.
\newblock {\em JHEP} {\bf 2024}, {\em 2},~39.  
\newblock {\url{https://doi.org/10.1007/JHEP02(2024)039}}.

\bibitem[Hell and Lust(2025)]{Hell:2025wha}
Hell, A.; Lust, D.
\newblock {Conformal and pure scale-invariant gravities in d dimensions}.
\newblock {\em JHEP} {\bf 2025}, {\em 9},~202.  
\newblock {\url{https://doi.org/10.1007/JHEP09(2025)202}}.

\bibitem[Hell et~al.(2026)Hell, Ferreira, Lust, and Sasaki]{Hell:2026blj}
Hell, A.; Ferreira, E.G.M.; Lust, D.; Sasaki, M.
\newblock {The recipe for the degrees of freedom}. \emph{arXiv} {\bf 2026}, 	arXiv:2601.10288.

\bibitem[Starobinsky and Schmidt(1987)]{starobinsky1987general}
Starobinsky, A.; Schmidt, H.J.
\newblock On a general vacuum solution of fourth-order gravity.
\newblock {\em Class. Quantum Gravity} {\bf 1987}, {\em 4},~695.

\bibitem[Barrow and Ottewill(1983)]{barrow1983stability}
Barrow, J.D.; Ottewill, A.C.
\newblock The stability of general relativistic cosmological theory.
\newblock {\em J. Phys. A Math. Gen.} {\bf 1983}, {\em 16},~2757.

\bibitem[Hu and Sawicki(2007)]{hu2007models}
Hu, W.; Sawicki, I.
\newblock Models of f (R) cosmic acceleration that evade solar system tests.
\newblock {\em Phys. Rev. D—Part. Fields Gravit. Cosmol.} {\bf 2007}, {\em 76},~064004.

\bibitem[Riess et~al.(2022)Riess, Yuan, Macri, Scolnic, Brout, Casertano, Jones, Murakami, Anand, Breuval et~al.]{Riess2022}
Riess, A.G.; Yuan, W.; Macri, L.M.; Scolnic, D.; Brout, D.; Casertano, S.; Jones, D.O.; Murakami, Y.; Anand, G.S.; Breuval, L.; et~al.
\newblock A Comprehensive Measurement of the Local Value of the Hubble Constant with 1 km s$^{-1}$ Mpc$^{-1}$ Uncertainty from the Hubble Space Telescope and the SH0ES Team.
\newblock {\em Astrophys. J. Lett.} {\bf 2022}, {\em 934}, L7.

\bibitem[Collaboration(2018)]{Planck2018}
Collaboration, P.
\newblock Planck 2018 results. VI. Cosmological parameters.
\newblock {\em A\&A} {\bf 2018}, \emph{641}, A6.

\bibitem[Di~Valentino et~al.(2021)Di~Valentino, Mena, Pan, Visinelli, Yang, Melchiorri, Mota, Riess, and Silk]{di2021realm}
Di~Valentino, E.; Mena, O.; Pan, S.; Visinelli, L.; Yang, W.; Melchiorri, A.; Mota, D.F.; Riess, A.G.; Silk, J.
\newblock In the realm of the Hubble tension—A review of solutions.
\newblock {\em Class. Quantum Gravity} {\bf 2021}, {\em 38},~153001.

\bibitem[Abdalla et~al.(2022)]{Abdalla:2022yfr}
Abdalla, E.;  et~al.
\newblock {Cosmology intertwined: A review of the particle physics, astrophysics, and cosmology associated with the cosmological tensions and anomalies}.
\newblock {\em JHEAp} {\bf 2022}, {\em 34},~49--211. 
\newblock {\url{https://doi.org/10.1016/j.jheap.2022.04.002}}.

\bibitem[Montani et~al.(2025)Montani, De~Angelis, and Dainotti]{montani2025decay}
Montani, G.; De~Angelis, M.; Dainotti, M.G.
\newblock Decay of dark energy into dark matter in a metric f (R) gravity: Effective running Hubble constant.
\newblock {\em Phys. Dark Universe} {\bf 2025}, \emph{49}, 101969.

\bibitem[Montani et~al.(2024)Montani, Carlevaro, and De~Angelis]{montani2024modified}
Montani, G.; Carlevaro, N.; De~Angelis, M.
\newblock Modified gravity in the presence of matter creation: Scenario for the late Universe.
\newblock {\em Entropy} {\bf 2024}, {\em 26},~662.

\bibitem[Dainotti et~al.(2021)Dainotti, De~Simone, Schiavone, Montani, Rinaldi, and Lambiase]{dainotti2021hubble}
Dainotti, M.G.; De~Simone, B.; Schiavone, T.; Montani, G.; Rinaldi, E.; Lambiase, G.
\newblock On the Hubble constant tension in the SNe Ia Pantheon sample.
\newblock {\em  Astrophys. J.} {\bf 2021}, {\em 912},~150.

\bibitem[Dainotti et~al.(2022)]{Dainotti2022b}
Dainotti, M.G.; De Simone, B.; Schiavone, T.; Montani, G.; Rinaldi, E.; Lambiase, G.; Bogdan, M.; Ugale, S.
\newblock Updated analysis of Hubble constant evolution using expanded Pantheon-like samples.
\newblock {\em Galaxies} {\bf 2022}, {\em 10}, 24.
\newblock {\url{https://doi.org/10.3390/galaxies10010024}}.

\bibitem[Nojiri et~al.(2022)Nojiri, Odintsov, and Oikonomou]{Nojiri:2022ski}
Nojiri, S.; Odintsov, S.D.; Oikonomou, V.K.
\newblock {Integral F(R) gravity and saddle point condition as a remedy for the H0-tension}.
\newblock {\em Nucl. Phys. B} {\bf 2022}, {\em 980},~115850. 
\newblock {\url{https://doi.org/10.1016/j.nuclphysb.2022.115850}}.

\bibitem[Simone et~al.(2024)Simone, van Putten, Dainotti, and Lambiase]{desimone2024doubletcosmologicalmodelschallenge}
Simone, B.D.; van Putten, M.H.P.M.; Dainotti, M.G.; Lambiase, G.
\newblock A doublet of cosmological models to challenge the H0 tension in the Pantheon Supernovae Ia catalog. \emph{arXiv}  \textbf{2024}, 	arXiv:2411.05744.

\bibitem[Dainotti et~al.(2024)Dainotti, Lenart, Yengejeh, Chakraborty, Fraija, Di~Valentino, and Montani]{Dainotti:2024aha}
Dainotti, M.G.; Lenart, A.L.; Yengejeh, M.G.; Chakraborty, S.; Fraija, N.; Di~Valentino, E.; Montani, G.
\newblock {A new binning method to choose a standard set of Quasars}.
\newblock {\em Phys. Dark Univ.} {\bf 2024}, {\em 44},~101428. 
\newblock {\url{https://doi.org/10.1016/j.dark.2024.101428}}.

\bibitem[Li et~al.(2018)Li, Xue, Liu, Wu, Yang, Shao, Quan, Tan, Tu, Liu, et~al.]{li2018measurements}
Li, Q.; Xue, C.; Liu, J.P.; Wu, J.F.; Yang, S.Q.; Shao, C.G.; Quan, L.D.; Tan, W.H.; Tu, L.C.; Liu, Q.;  et~al.
\newblock Measurements of the gravitational constant using two independent methods.
\newblock {\em Nature} {\bf 2018}, {\em 560},~582--588.

\bibitem[Gundlach and Merkowitz(2000)]{gundlach2000measurement}
Gundlach, J.H.; Merkowitz, S.M.
\newblock Measurement of Newton's constant using a torsion balance with angular acceleration feedback.
\newblock {\em Phys. Rev. Lett.} {\bf 2000}, {\em 85},~2869.

\bibitem[Wang and Chen(2020)]{wang2020constraints}
Wang, K.; Chen, L.
\newblock Constraints on Newton’s constant from cosmological observations.
\newblock {\em  Eur. Phys. J. C} {\bf 2020}, {\em 80},~570.

\bibitem[Weinberg(2008)]{weinberg2008cosmology}
Weinberg, S.
\newblock {\em Cosmology}; OUP Oxford: Oxford, UK
,  2008.

\bibitem[Khoury and Weltman(2004{\natexlab{a}})]{khoury2004chameleon}
Khoury, J.; Weltman, A.
\newblock Chameleon cosmology.
\newblock {\em Phys. Rev. D} {\bf 2004}, {\em 69},~044026.

\bibitem[Khoury and Weltman(2004{\natexlab{b}})]{khoury2004chameleon2}
Khoury, J.; Weltman, A.
\newblock Chameleon fields: Awaiting surprises for tests of gravity in space.
\newblock {\em Phys. Rev. Lett.} {\bf 2004}, {\em 93},~171104.

\bibitem[Carroll(1998)]{carroll1998quintessence}
Carroll, S.M.
\newblock Quintessence and the rest of the world.
\newblock {\em arXiv} {\bf 1998}, 	arXiv:astro-ph/9806099.

\bibitem[Quiros et~al.(2015)Quiros, Garcia-Salcedo, Gonzalez, and Horta-Rangel]{quiros2015chameleon}
Quiros, I.; Garcia-Salcedo, R.; Gonzalez, T.; Horta-Rangel, F.A.
\newblock The Chameleon Effect in the Jordan Frame of the Brans--Dicke Theory.
\newblock {\em arXiv} {\bf 2015},   arXiv:1506.05420.

\bibitem[Brax et~al.(2008)Brax, van~de Bruck, Davis, and Shaw]{brax2008f}
Brax, P.; van~de Bruck, C.; Davis, A.C.; Shaw, D.J.
\newblock f (R) gravity and chameleon theories.
\newblock {\em Phys. Rev. D—Part. Fields Gravit. Cosmol.} {\bf 2008}, {\em 78},~104021.

\bibitem[Capozziello and Tsujikawa(2008)]{PhysRevD.77.107501}
Capozziello, S.; Tsujikawa, S.
\newblock Solar system and equivalence principle constraints on $f(R)$ gravity by the chameleon approach.
\newblock {\em Phys. Rev. D} {\bf 2008}, {\em 77},~107501.
\newblock {\url{https://doi.org/10.1103/PhysRevD.77.107501}}.

\bibitem[Liu et~al.(2018)Liu, Zhang, and Zhao]{LIU2018286}
Liu, T.; Zhang, X.; Zhao, W.
\newblock Constraining f(R) gravity in solar system, cosmology and binary pulsar systems.
\newblock {\em Phys. Lett. B} {\bf 2018}, {\em 777},~286--293.
\newblock {\url{https://doi.org/10.1016/j.physletb.2017.12.051}}.

\bibitem[Hernández-Aguayo et~al.(2018)Hernández-Aguayo, Baugh, and Li]{10.1093/mnras/sty1822}
Hernández-Aguayo, C.; Baugh, C.M.; Li, B. 
\newblock Marked clustering statistics in f(R) gravity cosmologies.
\newblock {\em Mon. Not. R. Astron. Soc.} {\bf 2018}, {\em 479},~4824--4835. 
\newblock {\url{https://doi.org/10.1093/mnras/sty1822}}.

\bibitem[Amendola et~al.(2007)Amendola, Gannouji, Polarski, and Tsujikawa]{amendola2007conditions}
Amendola, L.; Gannouji, R.; Polarski, D.; Tsujikawa, S.
\newblock Conditions for the cosmological viability of f (R) dark energy models.
\newblock {\em Phys. Rev. D—Part. Fields Gravit. Cosmol.} {\bf 2007}, {\em 75},~083504.


\bibitem[Mandl and Shaw(2013)]{mandl2013quantum}
Mandl, F.; Shaw, G.
\newblock {\em Quantum Field Theory}; John Wiley \& Sons:  Hoboken, NJ, USA, 
  2013.

\bibitem[Friedman(1922)]{friedman1922krummung}
Friedman, A.
\newblock {\"U}ber die kr{\"u}mmung des raumes.
\newblock {\em Z. Phys.} {\bf 1922}, {\em 10},~377--386.

\bibitem[Lema{\^\i}tre(1927)]{lemaitre1927univers}
Lema{\^\i}tre, G.
\newblock Un Univers homog{\`e}ne de masse constante et de rayon croissant rendant compte de la vitesse radiale des n{\'e}buleuses extra-galactiques.
\newblock {\em Ann. Soci{\'e}t{\'e} Sci.  Brux.} {\bf 1927}, {\em 47},~49--59.

\bibitem[Robertson(1935)]{robertson1935kinematics}
Robertson, H.P.
\newblock Kinematics and world-structure.
\newblock {\em Astrophys. J.} {\bf 1935}, {\em 82},~284.

\bibitem[Walker(1937)]{walker1937milne}
Walker, A.G.
\newblock On Milne's theory of world-structure.
\newblock {\em Proc. Lond. Math. Soc.} {\bf 1937}, {\em 2},~90--127.

\bibitem[Efstathiou and Gratton(2020)]{efstathiou2020evidence}
Efstathiou, G.; Gratton, S.
\newblock The evidence for a spatially flat Universe.
\newblock {\em Mon. Not. R. Astron. Soc. Lett.} {\bf 2020}, {\em 496},~L91--L95.

\bibitem[Di~Valentino et~al.(2025)Di~Valentino, Said, Riess, Pollo, Poulin, G{\'o}mez-Valent, Weltman, Palmese, Huang, van~de Bruck, et~al.]{di2025cosmoverse}
Di~Valentino, E.; Said, J.L.; Riess, A.; Pollo, A.; Poulin, V.; G{\'o}mez-Valent, A.; Weltman, A.; Palmese, A.; Huang, C.D.; van~de Bruck, C.;  et~al.
\newblock The CosmoVerse White Paper: Addressing observational tensions in cosmology with systematics and fundamental physics.
\newblock {\em Phys. Dark Universe} {\bf 2025}, {\em 49},~101965.

\bibitem[Navonea et~al.(2025)Navonea, Dainotti, Fazzari, Montani, and Maki]{navonea2025creation}
Navonea, I.; Dainotti, M.G.; Fazzari, E.; Montani, G.; Maki, N.
\newblock Creation of Viscous Dark Energy by the Hubble Flow: Comparison with SNe Ia Master Sample Binned Data.
\newblock {\em arXiv} {\bf 2025},   arXiv:2511.16130.

\bibitem[Fazzari et~al.(2025)Fazzari, Dainotti, Montani, and Melchiorri]{fazzari2025effective}
Fazzari, E.; Dainotti, M.; Montani, G.; Melchiorri, A.
\newblock The effective running Hubble constant in SNe Ia as a marker for the dark energy nature.
\newblock {\em arXiv} {\bf 2025},   arXiv:2506.04162.

\bibitem[Olmo(2007)]{olmo2007violation}
Olmo, G.J.
\newblock Violation of the equivalence principle in modified theories of gravity.
\newblock {\em Phys. Rev. Lett.} {\bf 2007}, {\em 98},~061101.

\bibitem[Farajollahi and Salehi(2010)]{farajollahi2010cosmic}
Farajollahi, H.; Salehi, A.
\newblock Cosmic dynamics in chameleon cosmology.
\newblock {\em Int. J. Mod. Phys. D} {\bf 2010}, {\em 19},~621--633.

\bibitem[Montani et~al.(2025)Montani, Escamilla, Carlevaro, and Di~Valentino]{montani2025decay2}
Montani, G.; Escamilla, L.A.; Carlevaro, N.; Di~Valentino, E.
\newblock Decay of $ f (R) $ quintessence into dark matter: Mitigating the Hubble tension?
\newblock {\em arXiv} {\bf 2025},   arXiv:2512.20193.

\bibitem[Kruglov(2023)]{kruglov2023logarithmic}
Kruglov, S.
\newblock Logarithmic gravity model.
\newblock {\em Int. J. Mod. Phys. D} {\bf 2023}, {\em 32},~2350037.

\bibitem[Lecian and Montani(2009)]{lecian2009implications}
Lecian, O.M.; Montani, G.
\newblock Implications of non-analytic gravity at solar system scales.
\newblock {\em Class. Quantum Gravity} {\bf 2009}, {\em 26},~045014.

\bibitem[Fiorucci et~al.(2014)Fiorucci, Lecian, and Montani]{fiorucci2014non}
Fiorucci, D.; Lecian, O.M.; Montani, G.
\newblock Non-analytical power-law correction to the Einstein--Hilbert action: Gravitational wave propagation.
\newblock {\em Mod. Phys. Lett. A} {\bf 2014}, {\em 29},~1450178.

\bibitem[Capozziello et~al.(2011)Capozziello, Carlevaro, De~Laurentis, Lattanzi, and Montani]{capozziello2011cosmological}
Capozziello, S.; Carlevaro, N.; De~Laurentis, M.; Lattanzi, M.; Montani, G.
\newblock Cosmological implications of a viable non-analytical f(R)-gravity model.
\newblock {\em arXiv} {\bf 2011},   arXiv:1104.2169.

\bibitem[Capozziello et~al.(2013)Capozziello, Carlevaro, De~Laurentis, Lattanzi, and Montani]{capozziello2013cosmological}
Capozziello, S.; Carlevaro, N.; De~Laurentis, M.; Lattanzi, M.; Montani, G.
\newblock Cosmological implications of a viable non-analytical f(R) model.
\newblock {\em  Eur. Phys. J. Plus} {\bf 2013}, {\em 128},~155.

\end{thebibliography}
\end{document}